\documentclass[11pt,letterpaper]{article}
\usepackage{latexsym}
\usepackage{epsfig}
\usepackage{graphicx}
\usepackage{amssymb}
\begin{document}

\newcommand {\eqn}[1] {eqn.~(\ref{#1})}
\newcommand {\fig}[1] {fig.~\ref{#1}}

\begin{center}
{\LARGE Force steps during viral DNA packaging ?} \\
\vspace{0.5cm}
\textsc{Prashant K. Purohit${}^*$, Jan\'{e} Kondev${}^\dag$,
and  Rob Phillips${}^*$ } \\
\vspace{0.25cm}
\emph{${}^*$Division of Engineering and Applied Science, California
Institute of Technology, Pasadena, CA 91125. \\
${}^\dag$Physics Department, Brandeis University, Waltham, MA 02454.} \\
\vspace{0.7cm}
\end{center}

\begin{abstract}
Biophysicists and structural biologists increasingly acknowledge the role 
played by the mechanical properties of macromolecules as a critical
element in many biological processes. This change has been brought about,
in part, by the advent of single molecule biophysics techniques that have made it
possible to exert piconewton forces on key macromolecules and observe
their deformations at nanometer length scales, as well as to observe the
mechanical action of macromolecules such as molecular motors.  This has
opened up immense
possibilities for a new generation of mechanical investigations that
will respond to such measurements in an attempt to develop
a coherent theory for the mechanical behavior of macromolecules
under conditions where thermal and chemical effects are on an equal
footing with deterministic forces. This paper presents an application
of the principles of mechanics to the problem of DNA packaging, one of the
key events in the life cycle of bacterial viruses with special reference
to the nature of the internal forces that are built up during the DNA 
packaging process.
\end{abstract}

\section{Introduction}
Mechanics has a long and rich tradition of reaching out to other fields.
For example, the mechanical analysis of thin films has revealed insights into
applications ranging from microelectronics to MEMS (see Freund and Suresh (2003)).
Mechanics arguments have demonstrated that dislocations and surface 
instabilities play a key role in determining the mechanical and 
electrical properties of such films. While the study of thin films
is a relatively new area to
have been influenced by mechanics there are others, like geology and
metallurgy, that have drawn on mechanics for  decades. The most recent
developments in intersonic crack propagation (Rosakis (2002), Rice (2001))
have direct
analogues in the study of motion along geological fault lines. Geophysicists
have relied heavily on data emerging from shock compression of solids to
model the thermomechanical behavior of planets (see Meyers (1994)). In addition, the impact of dislocation
mechanics, phase transformations and fracture mechanics on the science
of metallurgy can scarcely be overestimated.

Another field that promises to be visited by mechanicians with increasing
regularity is biology, though
the use of mechanical principles for both fluids and solids in understanding
biological processes is not at all new. For example, slender body theory from fluid
mechanics has been fruitfully applied to the study of motion in bacteria
and other microorganisms (see Bray (2001)). In addition, variational principles 
for determining
optimal shapes have been used for the study of red-blood cells (see Boal (2002)).
Evans and Skalak (1980) show how lipid bilayer vesicles can be modeled as 
shells and membranes with no shear resistance. The flagella of bacteria have been
idealized as rods in an effort to study how their rotation can give rise
to propulsion (Goldstein {\it et al.} (1998)).
The cytoskeletal framework of the cell is studied as a three dimensional
network of elements to obtain estimates of the response of cells to external
forces (see Boal (2002)). In this paper we restrict our attention to problems
in biological nanomechanics with the observation that there are a range of 
fascinating connections between mechanics and biology at larger scales as well.
One of the most intriguiging features of problems in biological nanomechanics 
is that they involve a rich interplay of statistical and deterministic forces 
as will be seen below.

It is only recently, however, that structural biologists have begun to understand
the significance of mechanical properties for the macromolecular processes that
sustain life. The role of bending stiffness of DNA in gene regulation
(Widom (2001)) and the stiffness of substrate in cell migration
(Lo {\it et al.} (2000)) are two examples. Such
insights have been garnered with the aid of sophisticated experiments that
have made it possible to apply piconewton forces on nanometer size objects
and measure the resulting deformations. The data from these experiments point
to a rich interplay of thermal and chemical forces with electrical and
mechanical forces. In this paper we show how experimental insights can be
combined with mechanical principles to construct a simple and quantitative 
mechanical theory of DNA packaging in bacterial viruses that suggests 
a new round of experiments.

\section{Mechanics and the Viral Life Cycle}
Viruses have been studied extensively (see Alberts {\it et al.} (1997))
in the last three decades not just
with the goal of understanding their pathogenic nature but also for gaining
insights into structural biology of proteins and nucleic acids and for
studying gene regulation. Even though viruses are perhaps the simplest entities from
a biological perspective, they are nanotechnological marvels that embody a
wealth of physics at the nanometer scale. The principles operating at these scales
are being rapidly unravelled through ingenious experiments, such as that
by Smith {\it et al.} (2001) on DNA packaging in the \(\phi\)29 virus, which
serves as the primary motivation for the theoretical analysis presented below.

\begin{figure}[h]
 \centering
 \includegraphics[scale=0.5]{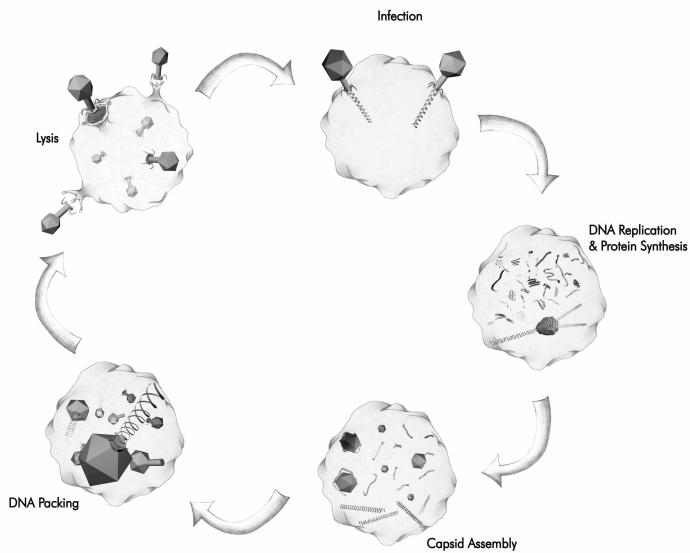}
 \caption{The life cycle of a virus. The critical
 step which is the focus of this paper is the packaging of DNA into the
 assembled capsid. The packaging process involves large forces on the
 order of several tens of pico-newtons and it proceeds at high rates,
 packing the full length of the genome in only a few minutes.}
 \label{fig:lcycle}
\end{figure}

Before going into the details of the experiment it is useful to get a
glimpse of the life-cycle of a virus (see fig.~\ref{fig:lcycle}). We are 
especially interested in the class of 
viruses known as bacteriophage and which infect bacteria
such as the well-known {\it E.coli}. In simplest terms, a bacteriophage
is nothing but a protein coat, the {\it capsid}, filled with nucleic acid 
such as DNA or RNA. The capsids of bacteriophage attach to the surface of
the bacterial
cell which is under attack. The genetic material of the virus is {\it ejected} into
the bacterium leaving the capsid behind. Though the process 
of DNA ejection is not completely
understood, it is argued that a mature bacteriophage capsid is highly
pressurised (see Smith {\it et al.} (2001)) and the genome is forcefully released
into the host cell whose contents are effectively at a much lower pressure.

Once the viral DNA is inside the bacterium it hijacks the protein production  
machinery of the bacterium to synthesize its own proteins such as those that
will ultimately make up the capsid. Copies of the viral DNA are also made.
Here again, experiments have shed light on the mechanics and kinetics of the
processes of {\it transcription} of the genome and its {\it translation}
into proteins. The process of transcription is mediated by a large protein
called RNA polymerase. This protein attaches to the DNA and then moves
along its length transcribing the DNA into
a molecule known as mRNA. As it moves along, the protein exerts forces
on the DNA molecule. These forces have been measured experimentally
(Wang {\it et al.} (1998)) and it
has been found that the protein can be stalled by exerting a force of roughly
25pN. In fact, the rate of transcription is heavily dependent on the force
exerted. The type of data emerging from such experiments unmistakably points
to a plethora of mechanics problems still to be resolved at the level of
macromolecular assemblies.

After a sufficient quantity of the viral proteins have been synthesized they
begin to {\it self-assemble} into hollow capsids. In the case of the \(\phi\)29
virus, the capsids have a portal at one end with an attached protein motor.  
The motor
proteins identify the appropriate end of the as-yet unpackaged viral
DNA and  push it into
the capsid. As more  of the DNA gets pushed into the capsid the motor
has to perform work against an increasing resistive force due to confinement of
the packaged DNA. This causes the motor to
slow down as more DNA fills up, but ultimately all the DNA is packed
 and the remaining proteins attach themselves to the capsid thus making it
ready to infect another bacterium. Once  the capsids are fully assembled an
enzyme is released that breaks up the bacterial cell membrane and releases
the mature viruses so that they can repeat their evil action elsewhere (see Ptashne (1992)).

As noted above, a critical step in the life cycle (see fig.~\ref{fig:lcycle})
of the virus is
the packaging of DNA into the capsid by the portal motor. This step was the
focus of the experiment by Smith {\it et al.} (2001) in which they used optical
tweezers to measure the force exerted by the portal motor of the \(\phi\)29
bacteriophage on the DNA as a function of the amount of DNA already packed in
the capsid. One end of
the DNA was attached to a silica bead which was held in an optical trap while the
motor tugged on the other end. The capsid itself was  attached to
a second silica bead that was held by a glass pipette thus immobilizing the
capsid during the packaging process. Through a series of careful
experiments, the average force exerted by the portal motor was measured and
plotted against the fraction of packed DNA. The rate at which the DNA is
packed was also measured as a function of the fraction packed. It was 
found that during the early stages of packing the motor packs at a rate of
roughly 100 base pairs per second which correponds to \(34\)nm per second. 
However, the
rate drops considerably after about \(70\)\% of the DNA has been packed and   
it slows  to a crawl as the packaging nears completion. We note
that the experiment was conducted {\it in vitro} in a solution which contained
\(50\)mM NaCl, \(5\)mM MgCl\(_{2}\) and tris-HCl buffer at pH 7.8. These
are not the conditions prevailing in a bacterium and one of the
interesting outcomes of the type of model being described here is
the recognition that the ionic strength is a key control parameter in
governing the physics of DNA packaging. 
There are a variety of interesting mechanics issues to be confronted
relating to those stages of the viral life-cycle involving
both DNA packaging and ejection. One such issue involves determining what
is the maximum sustainable
pressure in a viral capsid (Purohit {\it et al.} (2003)). Another concerns
determining the amount
of DNA ejected by a bacteriophage as a function of the osmotic pressure in the
exterior solution (Evilevitch {\it et al.} (2003)).

This paper puts forth a simple mechanical model of the packaging process in an
endeavour to understand the nature of the forces being exerted on the DNA.
Previously, we described results of an analysis based on a continuum
approximation to the mechanical model of the packaging force (Purohit {\it et al.}
(2003)). Here we analyze a discrete version of that same model which 
reveals interesting new effects which could be accessible experimentally.

\section{The Free Energy Function for Packed DNA}
In order to estimate the energetics of viral DNA packing, 
we take a structural hint from the cryo-electron microscopy images of capsids
(see, for example, Cerritelli {\it et al.} (1997)) in various stages of the
packing process. It has been observed for several different
viruses that the packed DNA is arranged  in a series of circular hoops (or
a spool). Each hoop finds itself at the center of a hexagon formed
by nearly parallel segments of the same overall strand.  The exception are 
the hoops 
hugging the surface of the capsid, or those at the inner surfaces of the
spool, which are
surrounded on average by three nearest neighbors. We note that the geometry is
more subtle than the hoop packing adopted here since the packing involves a 
helical pitch and hence the adjacent strands are probably not perfectly
parallel. Following earlier
work (see Riemer and Bloomfield (1978), Odijk (1998) and Kindt {\it et al.}
(2001)), the basic idea is to write a free energy function which characterizes
the energetics of structures like those described above as a sum of elastic
terms and interaction terms.  It is well known that DNA
has a considerable bending stiffness at length scales of a few tens of
nanometers. This is evident from the fact that the persistence length of DNA is
about \(50\)nm. The persistence length is defined as the distance over which the
tangents to a fiber are correlated. The more flexible a fiber is the smaller its
persistence length. Polymer physicists use the persistence length as a measure
of the stiffness of a chain (see Doi and Edwards (1988)). Since the

 persistence length of DNA is of the same order
as the size of viral capsids 
we expect the {\it bending} energy to play an important part in the packaging
process.
Such a description of the elasticity of DNA is now quite firmly established in
literature and is found to be useful in understanding key biological
processes related to genetic regulation and DNA condensation in chromosomes
of eucaryotic cells.

Beyond its explicit mechanical properties another relevant characteristic of DNA is
its strong acidity. In particular, in an aqueous solution the backbone
of the molecule is highly charged, with two units of negative charge per 
base pair. As a result there
are electrostatic penalties involved in bringing two strands in close
proximity. In the salty solutions found in cells the picture is more complicated 
because charged ions screen these interactions. Another complication arises 
from the 
fact that close packing involves the ejection of water molecules from the 
vicinity of the strands, which make sizable contributions to the free energy.
All of these complex interactions of the DNA with the ambient solution and with
itself result in an effective {\it interaction} energy. The
importance of these effects is immediately evident when we look at the physics
of DNA condensation in a solutions containing trivalent and tetravalent cations.
In particular, DNA is known to form hexagonally packed toroidal structures when
immersed in a solution containing these ions (Raspaud {\it et al.} (1998)).
The aim of the calculations to follow is to model viral packing, taking into account 
both elastic and interaction energies. 

\subsection{Elastic Contribution to the Free Energy}

We begin by idealizing DNA as an elastic rod 
 capable of sustaining bending deformations. The centerline of the
rod is parametrized by the reference coordinate \(x\). The bending energy stored
in length \(l\) of this rod is given by
\begin{equation}
 e(l) = \frac{\kappa}{2}\int_{0}^{l}\frac{dx}{R(x)^{2}},
\end{equation}
where \(R(x)\) is the radius of curvature at reference position \(x\) along
the rod and \(\kappa\) is a bending modulus which we assume to be independent
of the reference position \(x\) though we note that the

sequence dependence of the elasticity of DNA is one of its most

intriguing features. For a rod bent into a hoop, the
radius of curvature \(R(x)\) is just the radius of the hoop \(R\),  
and the length is \(l=2\pi R\), and so the integral reduces to
\begin{equation}
\label{eq:hoopel}
 e(2 \pi R) = \frac{\kappa}{2R^{2}}\int_{0}^{2\pi R}dx = \frac{\pi\kappa}{R}.
\end{equation}
The modulus \(\kappa\) is usually taken to be \(EI\) where \(E\) is the
Young's modulus of the material and \(I\) is the moment of inertia of the
cross-section. However, it is convenient to express the elastic properties 
of molecules differently since thermal oscillations play a dominant
role in their deformation. In particular, the deformation at a 
given point on a molecule may be  uncorrelated with the deformation at 
another location because of the randomness caused by these thermal
motions. As noted above, the length at which the importance of bending energy
is comparable to that due to thermal vibrations is called the `bend persistence 
length'
and is denoted by \(\xi_{p}\). More precisely, \(\xi_{p}\) is determined
by balancing the bending energy and the thermal energy and
results in  \(\kappa = \xi_{p}k_{B}T\),
where \(\xi_{p}\) is the
persistence length, \(T\) is the temperature and \(k_{B}\) is Boltzmann's
constant; \(\xi_{p} = 50\)nm for DNA and \(k_{B}T = 4.1\)pNnm at \(T=300\)K.
Using this description of the elastic properties of DNA and adding up
the contributions due to all of the hoops in the spool, the 
total bending energy is given by
\begin{equation}
 E_{bend} = \pi \xi_{p}k_{B}T\sum_{i}\frac{N(R_{i})}{R_{i}}, 
\end{equation}
where \(N(R_{i})\) is the number of hoops in a column of radius \(R_{i}\).
Note that we neglect the fact that the DNA is packed in a helical arrangement
with an associated contribution to the curvature due to the helical pitch.
As noted in our earlier paper (Purohit {\it et al.} (2003)), this effect is
negligible.
For analytical simplicity the discrete expression given above can be converted 
into an integral and written as
\begin{equation} \label{eq:engint}
 E_{bend} = \frac{2\pi \xi_{p}k_{B}T}{\sqrt{3}d_{s}}\int_{R}^{R_{out}}
            \frac{N(r)}{r}dr
\end{equation}
where \(d_{s}\) is the spacing between adjacent hoops, \(R_{out}\) is the
radius of the capsid and \(R\) is the radius of the innermost set of hoops.
The factor of \(\frac{\sqrt{3}}{2}\) appears since the spacing between two
adjacent columns of hoops is \(\frac{\sqrt{3}}{2}d_{s}\) in a hexagonal
array. The integral approximation has the virtue that it leads to 
closed-form analytical expressions for the energies and forces in the viral
packing problem,  at least for simple geometries like the cylinder and the sphere.
An interesting set of conclusions to be discussed below
are the differences between the discrete and continuum descriptions of this
problem, and their possible experimental consequences.

\subsection{Free Energy of Interaction}
In addition to the role played by elastic bending we must also consider the
effect of the interaction between adjacent strands of DNA. For this we appeal
to the experiments of Parsegian {\it et al.} (1986) and Rau {\it et al.} (1984).
In these experiments, osmotic pressure was applied on hexagonally packed
DNA and the interstrand separation was measured as a function of the
pressure for a variety of different solvent conditions. For conditions
comparable to those of Smith {\it et al.} (2001), it was found that the
osmotic pressure could be related to the interstrand spacing as
\begin{equation}
 p(d_{s}) = F_{0}\exp(-\frac{d_{s}}{c}),
\end{equation}
where \(F_{0}\) is a constant whose magnitude depends on the type and strength
of the ionic solution, and \(c= 0.27\)nm is a decay length which is roughly constant
over a  range of ionic conditions. For a solution containing
\(500\)mM NaCl at 298K, measurements reveal
\(F_{0}=55000\)pN/nm\(^{2}\). To this piece of empirical
evidence we add the assumption that parallel strands interact through a pair
potential \(v(d_{s})\) per unit length and that interactions are limited only
to  nearest neighbors. Given this description of the total energy
and the measured pressure vs interstrand spacing, we can deduce the functional
form of \(v(d_{s})\).

In particular, consider \(N\) parallel strands of length \(l\) each
packed in a hexagonal array with a spacing \(d_{s}\). The total interaction
energy of this arrangement is
\begin{equation} \label{eq:bass}
 E = 3N l v(d_{s}),
\end{equation}
where the factor of \(3\) appears because each strand interacts with \(6\) nearest
neighbours (ignoring surface effects) and we multiply by \(1/2\) to account for
double counting. The total volume of the assemblage is obtained by adding
together the (prismatic) volume occupied by each strand, thus
\begin{equation}
 V = N \frac{\sqrt{3}}{2} d_{s}^{2}l.
\end{equation}
We now use the thermodynamic identity \(p=-\frac{\partial E}{\partial V}\) to 
relate the expression
for pressure obtained from experiments to our simple model. We note that
\(dE = 3Nl \frac{\partial v}{\partial d_{s}} dd_{s}\) and 
\(dV = \sqrt{3} d_{s} Nl dd_{s}\), so that
\begin{equation}
 f(d_{s})= -\frac{\partial v(d_{s})}{\partial d_{s}} 
         = \frac{1}{\sqrt{3}} p(d_{s}) d_{s},
\end{equation}
where \(f(d_{s})\) is the force per unit length on the strands. We now
substitute the experimental result, namely 
\(p(d_{s}) = F_{0}\exp(-\frac{d_{s}}{c})\) and solve the
differential equation above with the boundary condition \(v(\infty)=0\). This
results in the potential
\begin{equation}
 v(d_{s}) = \frac{1}{\sqrt{3}}F_{0}(c^{2}+cd_{s}) \exp(-\frac{d_{s}}{c}).
\end{equation}
If this result is now exploited in the context of eqn. (\ref{eq:bass}) the 
interaction energy can be written as 
\begin{equation}
 E_{int} = \sqrt{3}F_{0}(c^{2}+cd_{s})L \exp(-\frac{d_{s}}{c}),
\end{equation}
where \(L=Nl\) is the total length of the strands.

Now we are in a position to write the total energy of the packaged DNA 
as a sum of the bending energy and the interaction energy as follows
\begin{equation} \label{eq:mona}
 E(L,d_{s}) = \sqrt{3}F_{0}L(c^{2}+cd_{s})\exp(-\frac{d_{s}}{c})
             + \pi\xi_{p}k_{B}T\sum_{i}\frac{N(R_{i})}{R_{i}}.
\end{equation}
In the continuum approximation this takes the form
\begin{equation} \label{eq:totint}
 E(L,d_{s}) = \sqrt{3}F_{0}L(c^{2}+cd_{s})\exp(-\frac{d_{s}}{c})
             +\frac{2\pi \xi_{p}k_{B}T}{\sqrt{3}d_{s}}\int_{R}^{R_{out}}
              \frac{N(r)}{r}dr.
\end{equation}
We still need to express the length \(L\) in terms of the radii of the hoops.
To do so, we simply add up the accumulated length of all the hoops as,
\begin{equation} \label{eq:discrlen}
 L = \sum_{i} 2\pi R_{i} N(R_{i}).
\end{equation}
In the continuum approximation this can be rewritten as
\begin{equation} \label{eq:lenint}
 L = \frac{4\pi}{\sqrt{3}d_{s}}\int_{R}^{R_{out}}rN(r)dr.
\end{equation}
With both  the elastic and interaction contribution to the total
energy of packed DNA in hand, we turn now to a concrete investigation of the
implications of this model.

\section{Geometry and Energetics of Packed DNA}
The calculation of the free energy associated with the DNA packed in a partially
filled capsid is predicated on key insights gained from experiments. We
assume throughout that the DNA adopts an inverse-spool geometry like
that described above, with the proviso that the class
of minimum free energy structures we consider is constrained to
the spool-type and that there are perhaps more complex
structures with lower free energy. We also assume that the strands have 
local six-fold coordination.
As a result, the only variable that remains to be determined is the spacing
\(d_{s}\) between the strands as a function of the amount of DNA packed.
In the regime where the interaction between the DNA strands is repulsive 
the interaction term dictates that the strands be as far apart as possible. 
On the other hand, since the outer
dimensions of the capsid are fixed, larger spacings imply smaller radii of
curvature which results in a steep rise in the bending energy cost.
The competition between these two contributions to the free energy determines
the geometry of the packed DNA (Reimer and Bloomfield (1978), Odijk (1998),
Kindt {\it et al.} (2001)). We illustrate this through two
examples and also make a comparison with the experimental data of
Smith {\it et al.} (2001).

\begin{figure}[h] 
 \centering
 \includegraphics[scale=0.6]{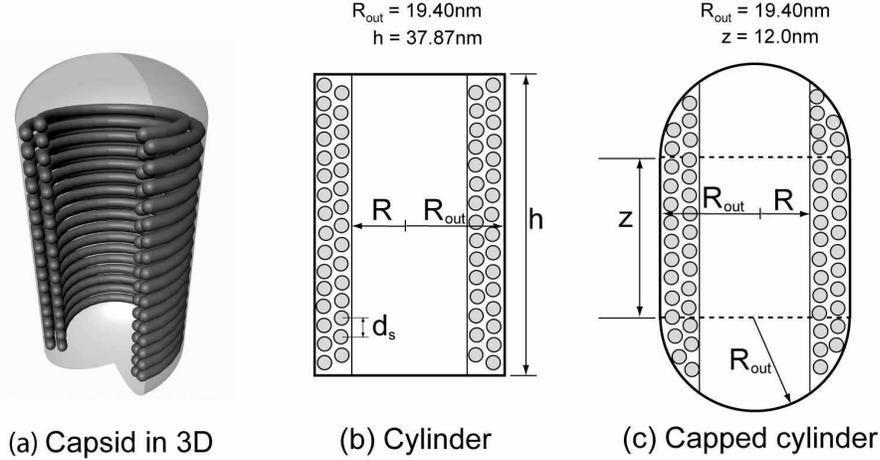}
 \caption{Model representations of viral capsids. (a) Capsid modeled as
 a cylinder with hemispherical caps shown in perspective. The curved rods
 inside represent the packed DNA. (b) The cylindrical geometry shown in 
 cross-section with height adjusted to correspond to the correct internal 
 volume of the capsid. The gray circles represent the hexagonally packed 
 DNA hoops. (c) Idealization of \(\phi\)29 as a cylinder with
 hemispherical caps shown in cross-section.}
\label{fig:phi29}
\end{figure}

For simplicity we begin by considering a cylindrical capsid which is a 
first approximation
to the geometry of the \(\phi\)29 virus considered in the experiments of Smith
{\it et al.} (2001). In reality, the \(\phi\)29 virus is shaped like a hollow 
oblong spheroid (Tao {\it et al.} (1998) and Wikoff and Johnson (1999)) 
which has an outer radius of
\(21\)nm and a height of about \(54\)nm. Its protein coat is about \(1.6\)nm
thick (on average) so that the radius of the capsid volume available to the DNA
is \(R_{out}=19.4\)nm while its height is roughly \(51.0\)nm. The capsid has
other interesting geometrical features, but for the purposes of the 
present calculation we idealize it as a cylinder. The dimensions of this 
idealized cylinder are given in fig.~\ref{fig:phi29} where the key point is
selecting the height of the cylinder so as to gaurantee that the volume
available for packing is identical to that of the real virus. The figure 
shows a second geometry that will be used later as a more refined model
of the \(\phi\)29 capsid.

The \(\phi\)29 capsid encloses about \(6.6\mu\)m of double-stranded DNA
corresponding to a genome of \(19.3\) kilobase pairs. As noted above for 
our cylindrical model we take the height of the cylinder as \(z=37.87\)nm 
to ensure that the internal volume is the same as that of the actual capsid.
For such a capsid, the number of hoops at radius \(R_{i}\) is given by
\(N(R_{i}) = h/d_{s}\) where \(d_{s}\) is the
separation between the strands and \(h\) is the height of the cylinder as shown
in fig.~\ref{fig:phi29}.
One of the key themes of the present paper is the comparison of the
discrete and continuous treatments of the structure and energetics
of viral DNA packing.  To that end, it is of interest to
examine the length of packaged DNA in the discrete setting. 
Assuming that there are \(n\) columns of stacked hoops (see
fig.~\ref{fig:phi29} where we show two completed columns of hoops, i.e.~\(n=2\)),
the length of the packed DNA is
\begin{eqnarray}
 L & =& 2\pi \frac{h}{d_{s}} \Big(R_{out} + R_{out}-\frac{\sqrt{3}}{2}d_{s}
       +R_{out} -2\frac{\sqrt{3}}{2}d_{s} + ...
       +R_{out}-(n-1)\frac{\sqrt{3}}{2}d_{s}\Big)  \nonumber \\
   & =& 2\pi n\frac{h}{d_{s}}\Big(R_{out}
       -\frac{n-1}{2}\frac{\sqrt{3}}{2}d_{s}\Big). \label{eq:dislen}
\end{eqnarray}
This equation results from adding up the accumulated length of the hoops at 
each radius as demanded by eqn.~(\ref{eq:discrlen}).
The hexagonal coordination of the strands appears once again through the
distance between adjacent columns of hoops which is  \(\frac{\sqrt{3}}{2}d_{s}\).
As a concrete example, we use the numbers for the \(\phi\)29 phage and 
compute how many
columns of stacked hoops it contains when fully packed. 
X-ray measurements on packed \(\phi\)29 indicate that the spacing between the
strands in a fully packed \(\phi\)29 capsid is about \(2.8\)nm (Earnshaw
and Casjens (1980)). The length of the genome is \(L=6584\) nm, and  
eqn. (\ref{eq:dislen}) implies \(n=5.61\) corresponding to \(5\) columns of
completely stacked
hoops and one innermost column of hoops that is only about \(60\%\) full.
There is a hollow cylindrical region inside whose radius is \(R_{in}
= 7.28\)nm. This calculation reveals that
there are not a large number of columns of hoops even in a completely
filled capsid and as a result, caution should be exercised in making the continuum
approximation in converting the discrete sums in equations (\ref{eq:engint})
and (\ref{eq:lenint}) to integrals.

We now turn to the more detailed aspects of the geometry and see how by 
minimizing the free energy we can obtain experimentally falsifiable insights 
into the packing process.

\subsection{Interstrand Spacing in Packed DNA Capsids: Discrete Model}
\begin{figure}[h]
 \centering
 \includegraphics[scale=0.6]{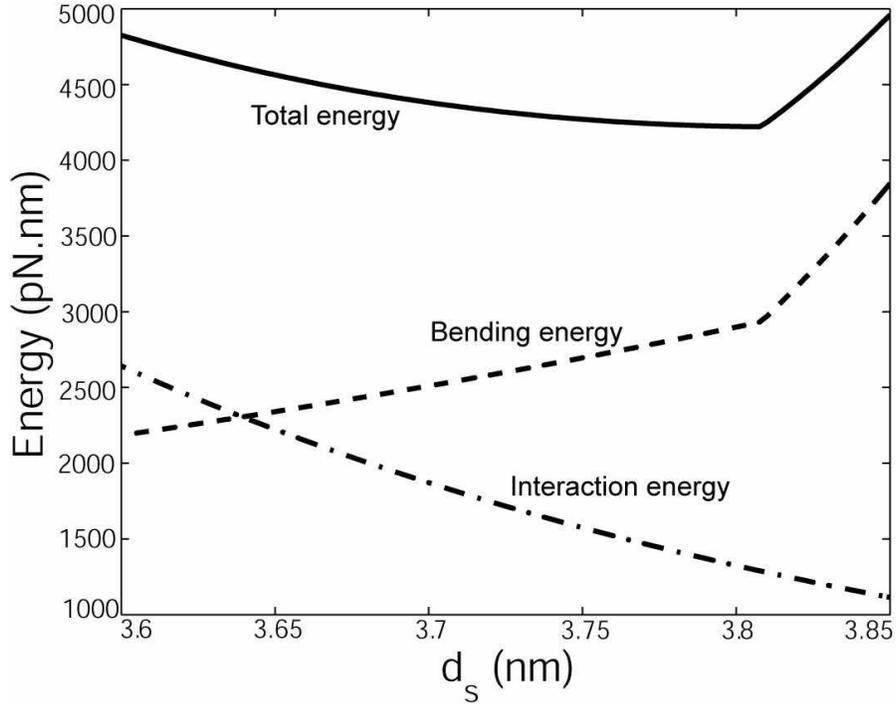}
 \caption{Energy vs interstrand spacing for \(L=4.0\mu\)m
 in a cylindrical capsid with \(R_{out}=19.4\)nm and \(z=37.87\)nm
 and with repulsive parameters
 \(F_{0} = 225500\)pN/nm\(^{2}\) and \(c=0.27\)nm.  A minimum is
 evident at \(d_{s} = 3.81\)nm. The sharp turn in the bending energy at
 \(d_{s} = 3.81\)nm is the point where a new column of hoops with smaller radius
 begins to form.}
 \label{engyeg}
\end{figure}

We mentioned earlier that the spacing between the strands is determined
by the competition between bending and interaction energies. We
demonstrate this minimization procedure with a very simple calculation. The model
as set forth above has one free parameter, namely, $F_0$, the parameter
characterizing the strength of the repulsion between adjacent DNA strands.  We fix
\(F_{0} = 225500\)pN/nm\(^{2}\) since, as shown below, this leads to the best
fit to the experiment of Smith {\it et al.} (2001). We note that the ion 
concentration in their experiment is small compared to the values reported 
by Rau {\it et al.} (1984), which implies that the repulsive forces
will be stronger than those measured by Rau {\it et al.} (1984)
resulting, in turn, in a larger value of \(F_{0}\). We continue with 
the cylindrical capsid geometry with the same internal volume as the
\(\phi\)29 phage.
The goal is to determine the spacing \(d_{s}\) between the strands  for a given
length \(L\) of genome packed. To that end, we minimize the free energy with
respect to \(d_{s}\) at fixed length \(L\). In particular, we set
\(d_{s}\) to a fixed value in eqn. (\ref{eq:dislen}) and then solve for 
the number of columns of hoops this implies while holding the length 
\(L\) fixed. We then have
all the information needed to calculate the energy using eqn. (\ref{eq:mona}).
We repeat these steps for a range of different values of \(d_{s}\) to get a plot 
of energy
vs \(d_{s}\). Finally, we use this plot to determine the \(d_{s}\)
corresponding to minimum energy at a given length of DNA. One such plot is
shown in fig.~\ref{engyeg}.
The interaction energy decreases as the spacing increases. This is expected since
the interaction energy is a decaying exponential and increasing $d_s$ implies
that the interactions come from deeper in the tail of the
pair potential. The bending energy
rises as the spacing increases since the radii of the innermost strands become
successively smaller. The total energy is dominated by
the interaction part at smaller spacings and by the bending energy at higher
spacings. For the case shown in fig.~\ref{engyeg} we can clearly identify 
a minimum corresponding to \(d_{s}=3.81\)nm
in the total energy.
\begin{figure}[h]
 \centering
 \includegraphics[scale=0.6]{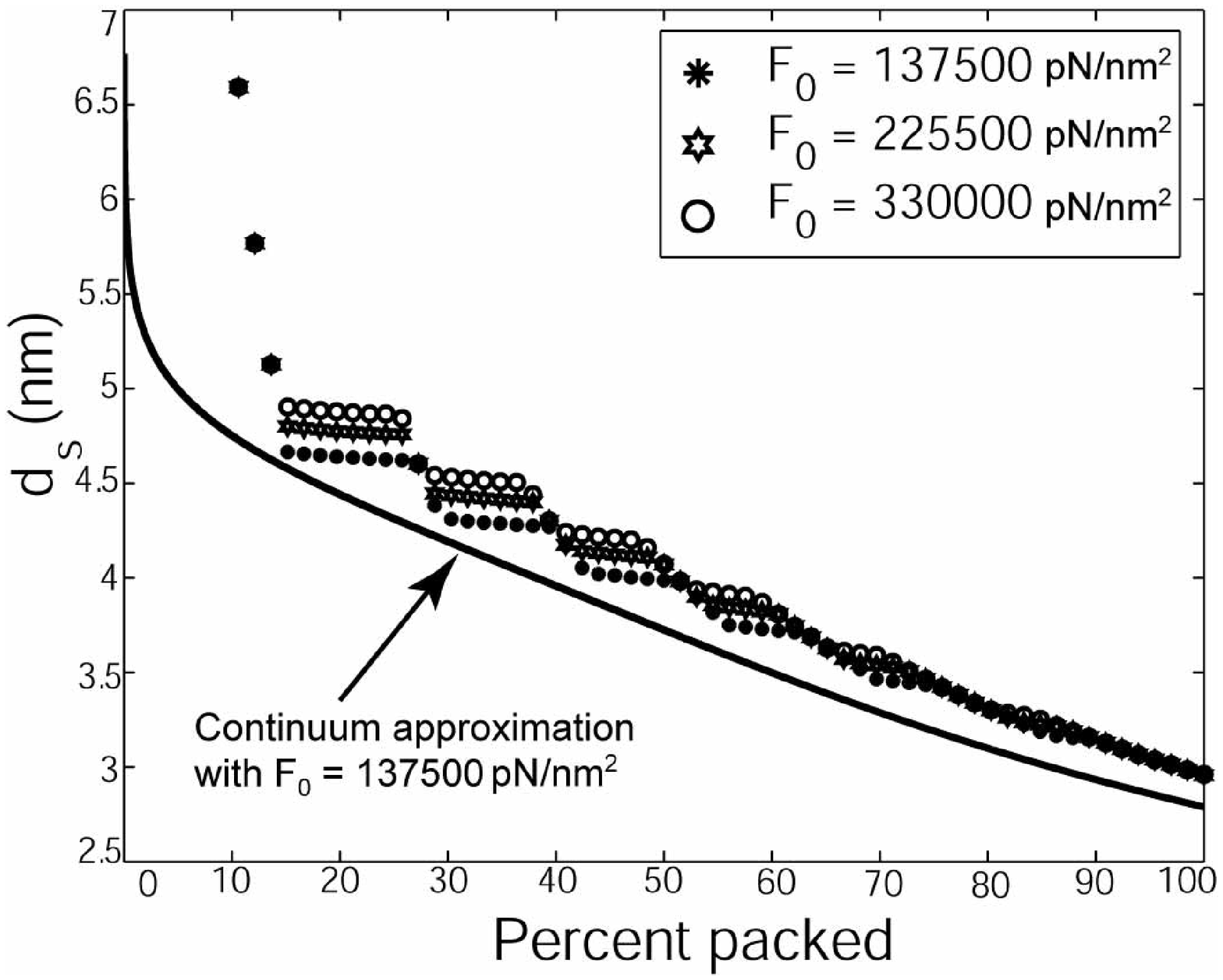}
 \caption{Spacing between the DNA strands as a function of the amount of DNA
 packed for a model cylindrical capsid. The spacing between strands in 
 the early stages is larger when
 the magnitude of the interaction energy is larger. In the later stages
 of packing the spacing is dictated by geometrical constraints. The points
 correspond to the discrete calculation for different ionic conditions and 
 the full curve corresponds to the continuum calculation.}
 \label{dsdisc}
\end{figure}
More generally to find the trends for all values of the length packed, 
this same procedure is repeated and then we
plot the optimal \(d_{s}\) (at minimum energy) against the fraction of DNA packed
for the \(\phi\)29 phage as shown in fig.~\ref{dsdisc} for
three different values of \(F_{0}\). We note that in the later stages of
packing all the curves collapse onto a single curve, though we also note that
the discrete and the continuous curves have different values in the large
packing limit. As shown in our earlier work (Purohit {\it et al.} (2003)), 
\(d_{s}\) in the large packing
limit is dictated by packing the DNA in such a way as to maximize the distance
between adjacent strands and should be seen as a geometric limit. 

On the basis of the simple model
presented above, fig.~\ref{dsdisc} provides several distinct predictions.
First, on a gross level,  it gives the history of the variation
of interstrand spacing during the packing process. This is a verifiable prediction
since it is possible to perform the viral packaging reaction for different ionic
concentrations of the ambient solution and for different values of the
genome length. Such experiments have been carried out for the T7 phage for 
three different
genome lengths by Cerritelli {\it et al.} (1997). Similar experiments have
also been performed for the \(\lambda\)-phage (see Earnshaw and Harrison (1977)).
The second more subtle outcome of the model is
the possibility that there are actually discrete effects in the packing
process due to the packing of the DNA at a finite set of discrete radii.
It would be of interest to determine whether these effects in the packing spacing
(and a related force signature to be described below) are present
in experiments.

\subsection{Interstrand Spacing in Packed Capsids: Continuum Model}
One of the interesting features of the calculations presented here is the
ability to contrast discrete and continuous descriptions of the energetics
of DNA packing.  To that end, we now repeat the energy minimization argument
that was used to determine the interstrand separation \(d_{s}\) in the 
discrete setting in the continuum approximation. In particular, we minimize 
the free energy given in eqn. (\ref{eq:totint}) with respect to \(d_{s}\), 
but subject to the constraint that \(L\) is constant. In concrete terms 
this amounts to 
\begin{equation}
 \frac{dE}{dd_{s}} = -d_{s}F_{0}L\exp(-\frac{d_{s}}{c})
 -\frac{4\pi\xi_{p}k_{B}T}{\sqrt{3}d_{s}^{3}}\int_{R}^{R_{out}}
  \frac{z(R')}{R'}dR' + \frac{4\pi\xi_{p}k_{B}T}{\sqrt{3}R^{2}d_{s}^{3}}
  \int_{R}^{R_{out}}R'z(R')dR'.
\end{equation}
where we have used the constraint \(\frac{dL}{dd_{s}} = 0\) which takes the
form
\begin{equation}
 \frac{dR}{dd_{s}} = -\frac{2}{d_{s}}\frac{\int_{R}^{R_{out}}R'z(R')dR'}{Rz(R)}.
\end{equation}
We substitute for \(L\) using \eqn{eq:lenint} and then set
\(\frac{dE}{dd_{s}}=0\) to get the following general equation
\begin{equation} \label{eq:gencont}
 \sqrt{3}F_{0}\exp(-\frac{d_{s}}{c}) = \frac{\xi_{p}k_{B}T}{R^{2}d_{s}^{2}}
 -\frac{\xi_{p}k_{B}T}{d_{s}^{2}}\frac{\int_{R}^{R_{out}}z(R')/R'dR'}
 {\int_{R}^{R_{out}}R'z(R')dR'}.
\end{equation}
This equation represents a competition between the interaction terms (on the
left) and the bending terms (on the right).
Note that the effect of the capsid geometry appears through the second term
on the right hand side of \eqn{eq:gencont}.
For a cylinder, \(z(R') = h\), a constant, and \eqn{eq:gencont} takes the form
\begin{equation} \label{eq:cylinnew}
 \sqrt{3}F_{0}\exp(-\frac{d_{s}}{c}) = \frac{\xi_{p}k_{B}T}{R^{2}d_{s}^{2}}
 -\frac{2\xi_{p}k_{B}T}{d_{s}^{2}}\frac{\log\big(\frac{R_{out}}{R}\big)}
  {R_{out}^{2}-R^{2}}
\end{equation}
%%\marginpar{\it RP: fix eqn. 19}
Eqn.~(\ref{eq:lenint}) is similarly specialized for a cylinder to yield
\begin{equation} 
\label{eq:cylingeo}
 L = \frac{2\pi h}{\sqrt{3}d_{s}^{2}}(R_{out}^{2}-R^{2}).
\end{equation}
Eqns.~(\ref{eq:cylinnew}) and (\ref{eq:cylingeo}) feature
two unknowns, \(R\) and \(d_{s}\), and we solve for them using the Newton-Raphson
method. The result of this computation is a history of the interstrand spacing
\(d_{s}\) as a function of the length \(L\) of DNA packed. This has been
plotted as a thick line in fig.~\ref{dsdisc} for
\(F_{0} = 137500\)pN/nm\(^{2}\). It is evident that while the continuum
version captures the trend quite well, it underestimates the value of
\(d_{s}\) for the cylindrical geometry in comparison with the more exact 
discrete version. Also, details captured by the discrete version, such as 
the series of steps and jumps, are smoothed out in the continuum 
approximation.

\section{Forces during DNA packaging}
One of the key outcomes of the model presented here is the internal force that 
builds up 
during DNA packing. The force vs percent packed curves should be seen as a 
second set of experimental implications for this mechanical model which 
complements predictions of the spacing \(d_{s}\) vs percent packed. 
\begin{figure}[h]
 \centering
 \includegraphics[scale=0.6]{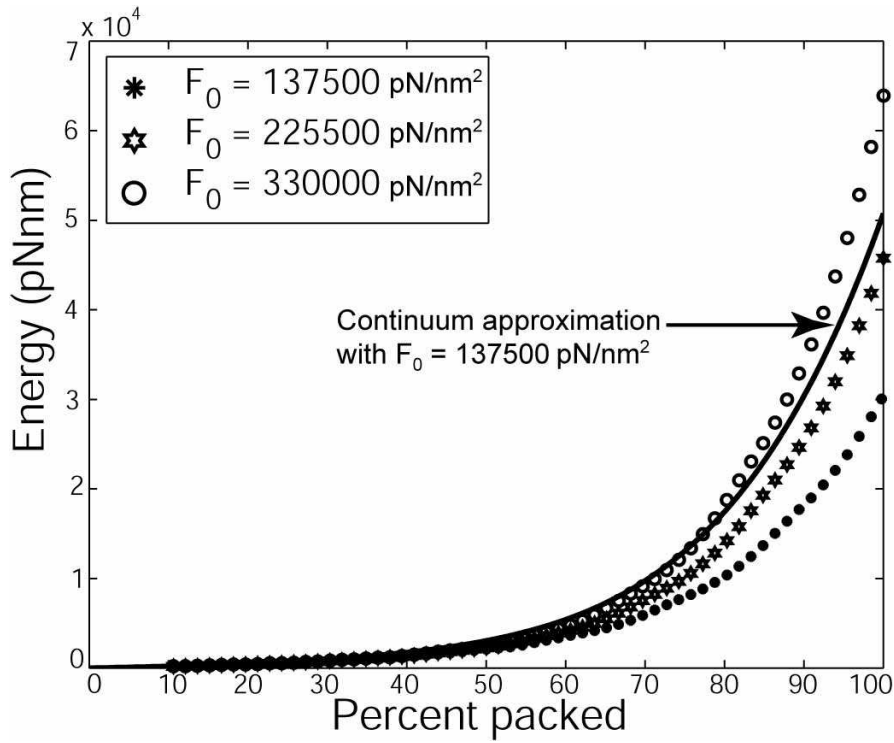}
 \caption{Variation in the stored energy as a function of the
 length of DNA packed for a model cylindrical capsid. The total energy rises 
 as a function of the amount of
 DNA packed and with increasing value of \(F_{0}\). Also shown as a thick
 line is the curve obtained from the continuum theory.  Because the continuum 
 version of $d_s$ is smaller than that obtained from the discrete model, there 
 is a corresponding increase in the interaction energy which causes the 
 continuum energy to be larger than its discrete counterpart.}
\label{fig:endrtcon}
\end{figure}
Indeed, once we have the spacing \(d_{s}\) as a function of the amount of 
DNA packed we can determine the energy using \eqn{eq:mona} or (\ref{eq:totint}) 
depending on whether we have adopted the discrete or continuum version of the 
free energy.  The variation
of the free energy has been plotted for the three \(F_{0}\) values in
fig.~\ref{fig:endrtcon}. Recall that the choice of \(F_{0}\) reflects the
ionic conditions present during the packaging process.
It is evident that the total energy increases at any given value of the
length packed as the strength of the repulsive interactions (the value
of \(F_{0}\)) is increased. Figure~\ref{fig:endrtcon} also shows the
energy obtained using the continuum approximation which overestimates the
energy but correctly captures the trend.
The force required to pack DNA is simply the derivative of the energy
with respect to the length packed. We have calculated this derivative
numerically and the results are plotted in fig.~\ref{fig:forcstep}.

The behavior of the force curves is very much as expected, with higher forces
for larger values of \(F_{0}\).   We note that the strong dependence of
the maximum packing force on $F_0$ is amenable to experimental 
observation and is worth further attention. In particular, we note that the 
maximum packing force can change by as much as a factor of two depending
upon the ionic conditions (and the related value of \(F_{0}\)). Perhaps 
the most striking feature of the force curves is the appearance of steps.
The same steps are seen in the plots of \(d_{s}\) as a function of fraction
packed. They occur when a new column (stack) of hoops starts to
form on the inside of the spool. Curiously enough, the experimental force curve
reproduced in fig.~\ref{coolfigr} also seems to show discontinuities of 
the slope at
4-5 different points, which might correspond to the five columns of completely
stacked hoops needed to fill the $\phi$29 capsid with its DNA.
However, to establish that this actually occurs
during the packaging of a virus would require a more detailed analysis of the
kinetics of packing and more careful experiments.  The presence
of such steps in the model results from implicitly assuming that the dynamics 
of packing follows the sort of organized column-by-column hoop packing 
described above. Therefore, the observation of such steps in experiment would
be not only a confirmation of the model, but a provocative hint about the 
dynamics of packing. 
\begin{figure}[h]
 \centering
 \includegraphics[scale=0.6]{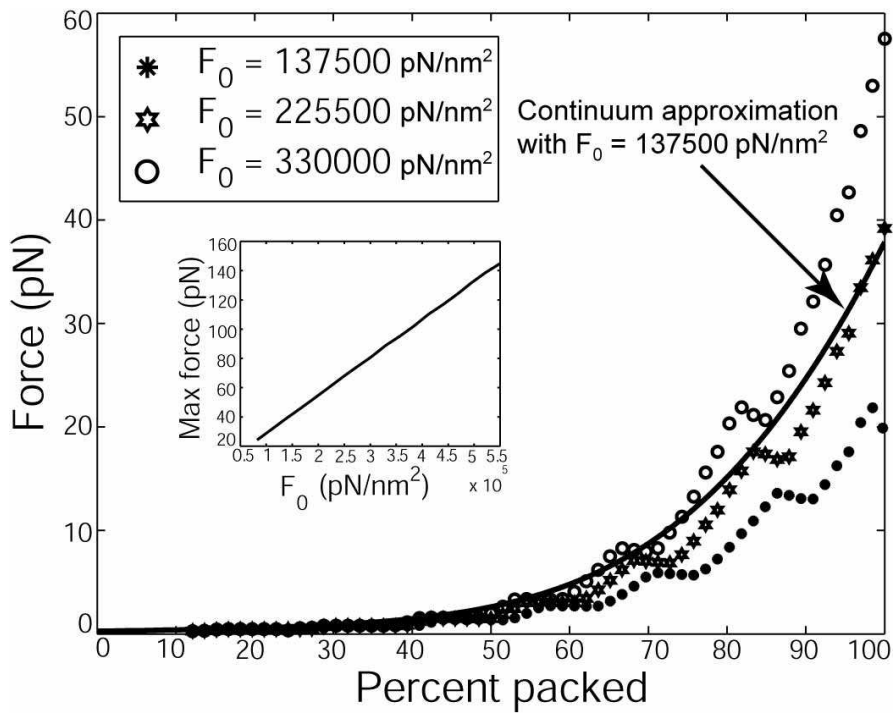}
 \caption{Internal force as a function of the amount of DNA packed for a model
 cylindrical capsid. The plot 
 bears the signature of both the total energy (\fig{fig:endrtcon}) and the 
 \(d_{s}\) (\fig{dsdisc}) curve. Note that the maximum force (at 100\% packing) 
 is roughly proportional to \(F_{0}\) as shown in the inset.}
 \label{fig:forcstep}
\end{figure}

\section{Application to the \(\phi\)29 virus}
\begin{figure}[h]
 \centering
 \includegraphics[scale=0.6]{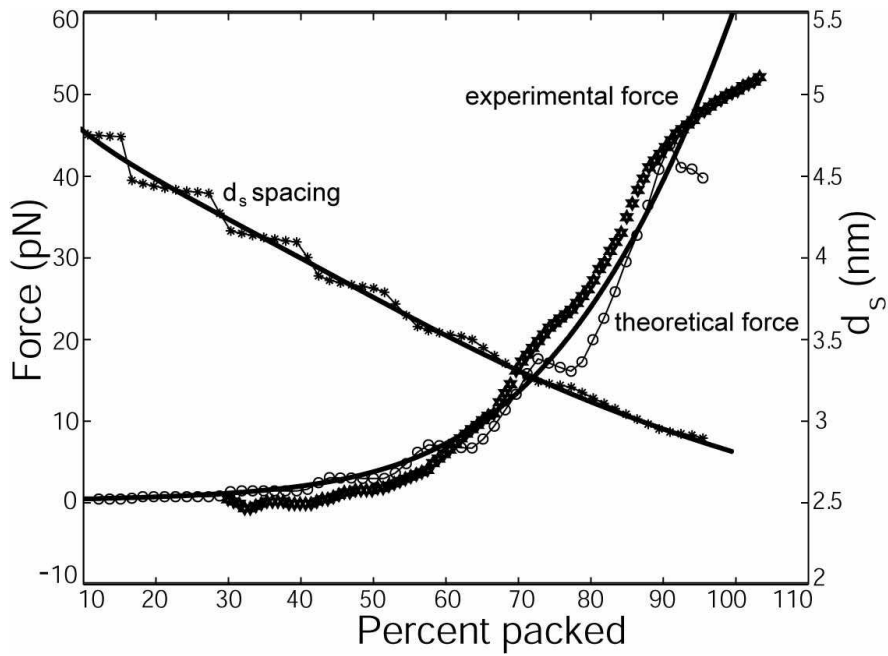}
 \caption{Force and interstrand spacing as functions of the amount of DNA
 packed in a capsid idealized as a cylinder with hemispherical caps. The 
 hexagons correspond to the experimental data of Smith {\it et al.} (2001).
 The thick lines are results of the continuum model and the circles connected 
 with the thin line are obtained from the discrete model. A good fit to 
 their data is obtained for \(F_{0} = 225500\) pN/nm\(^{2}\).}
 \label{coolfigr}
\end{figure}

As a final calculation we go beyond the analytic simplicity offered by the
cylindrical capsid and consider the second approximate geometry shown in 
fig.~\ref{fig:phi29} namely, the capped cylinder. 
Though the cylindrical geometry offers considerable insight into the 
mechanics of viral packaging it clearly represents an oversimplification 
of the \(\phi\)29 capsid.  The \(\phi\)29 capsid
is shaped like an oblong spheroid and we approximate it as a cylinder with
hemispherical caps in order to make comparisons with the experimental results
of Smith {\it et al.} (2001). The radius of the cylinder (and the hemispheres)
is taken to be \(R_{out}=19.4\)nm (see fig.~\ref{fig:phi29}). The height
of the cylindrical portion (also called the `waist') is \(z=12.0\)nm.
It is useful to study this geometry because one can obtain expressions for
spherical capsids (which are very good approximations to icosahedral viruses)
merely by setting the height of the waist to zero.
The expression for \(N(R_{i})\) is now a bit more complicated and is
given by
\begin{equation}
 N(R_{i}) = \frac{z + 2 \sqrt{R_{out}^{2} - R_{i}^{2}}}{d_{s}}.
\end{equation}
Using this formula in \eqn{eq:discrlen} results in the following expression for
the length of DNA packed:
\begin{eqnarray}
 L & =& 2\pi\Big[\frac{z}{d_{s}}R_{out}
              +\frac{z + 2 \sqrt{d_{s}(\sqrt{3}R_{out}
          -\frac{3}{4}d_{s})}}{d_{s}}(R_{out} - \frac{\sqrt{3}}{2}d_{s}) + ...
          \nonumber \\
   &  & +\frac{z + 2 \sqrt{(n-1)d_{s}(\sqrt{3}R_{out}
  -\frac{3}{4}(n-1)d_{s})}}{d_{s}}(R_{out} - (n-1)\frac{\sqrt{3}}{2}d_{s})\Big].
\end{eqnarray}
Unlike the cylindrical geometry, this expression is not analytically tractable. 
However, the same procedure can be used for determining the interstrand
spacing, and the energy and  force as a function of percent packed. 
The results of numerical calculations for this geometry, both in the discrete 
and the continuous setting, are plotted in fig.~\ref{coolfigr}.
We find that \(F_{0} = 225500\)pN/nm\(^{2}\) results in a good fit to the
experimental data. Moreover, the continuum and discrete versions of the theory
are in better agreement for this geometry than they were for cylindrical capsids.

It is of interest to examine the effect of ionic concentration on the forces
exerted by the portal motor on the DNA. The ionic concentration affects the
value of \(F_{0}\) and we have already seen that larger values of \(F_{0}\)
imply larger values of the force. In the inset to fig.~\ref{fig:forcstep}, the 
maximum force as a function of the parameter $F_0$ is shown.
This information can be potentially useful in the following way. It is known
(see Smith {\it et al.} (2001)) that the portal motor of the \(\phi\)29 virus
stalls at a force of \(57\)pN. In other words, if one were to conduct the
packaging reaction in a solution with ionic conditions such that 
\(F_{0}>250,000\)pN/nm\(^{2}\) then
the genome would not be completely packed since the motor would stall prior to
complete packing. We are hopeful
that experiments carried out with different ionic concentrations would
permit an investigation of such predictions.
We have made similar calculations to those presented here for viruses other than
\(\phi\)29 and find interesting variations in the maximum packing force from
one virus to the next which should be similarly accessible experimentally.

\section{Conclusion}
Recent advances in the development of tools for single molecule manipulation 
have permitted
the direct mechanical investigation of a variety of biological processes.
One intriguing recent example involves the direct measurement of mechanical
forces during DNA packaging in viruses. We have  developed
a simple, analytically tractable  model which responds to these experiments.
The model  emphasizes the role of elastic and electrostatic effects in
the packaging process and is in good  quantitative agreement with experimental data.
More importantly, the model makes a series of predictions that we hope will inspire
and guide future experiments.  In particular, we predict the appearance of 
steps in the force versus percent of DNA packed curve. There may be some 
evidence of steps in the existing experimental data, but further experimentation 
is needed.
The model also predicts a growth of the maximum force
required for packaging as the ion concentration is decreased; the underlying
cause being the stronger electrostatic repulsion between the DNA strands inside the
capsid. This points to
a new series of experiments that would look for incomplete packaging of
the viral DNA at lower salt concentrations than used previously. An alternative
approach would be to work at the same ion concentration but with longer
DNA constructs.

The biggest shortcoming of the model as presently stated is that it does not
address dynamical issues associated with the packaging process. The fundamental
experimental observation is that as the packaging process proceeds the packaging
rate falls from the initial 100 base pairs per second to zero. We note that 
the appearance of force steps described in this paper implicitly assumes a 
precise dynamical pathway resulting in a helical packing of the DNA in a series
of helices of ever decreasing radius and are hopeful that the future work will
shed further light on the dynamics of viral packing.

\section{Acknowledgements}
We are grateful to Ron Hockersmith, Paul Grayson, Dwight Anderson and Carlos
Bustamante for useful suggestions. We have benefitted from a long series
of stimulating discussions with Bill Gelbart, Alex Evilevitch and Chuck Knobler.
RP and PP acknowledge support of the NSF through grant number CMS-0301657, 
the NSF supported CIMMS center and the support of the Keck Foundation. JK 
is supported by the NSF under grant number DMR-9984471, and is a Cottrell 
Scholar of Research Corporation.


\begin{thebibliography}{99}

\bibitem{alberts1} Alberts B., Bray D., Johnson A., Lewis J., Raff M., Roberts K.
and Walter P., (1997) {\it Essential Cell Biology}, Garland Publishing, Inc.,
New York.

\bibitem{boal} Boal D., (2002) {\it Mechanics of the cell}, Cambridge University
press, Cambridge.

\bibitem{bray} Bray D. (2001) {\it Cell movements}, Garland Publishing, Inc.,
New York.

\bibitem{busta} Bustamante C., Macosko J.~C. and Wuite G.~J. (2000), Grabbing
the cat by the tail: Manipulating molecules one by one, {\it Nat. Rev. Mol.
Cell Biol.} {\bf 1}, 130-136.

\bibitem{cerritelli} Cerritelli, M.~E., Cheng, N., Rosenberg A.~H., McPherson C.~E.,
Booy, F.~P., and Alasdair C. Steven, A.~C., (1997), Encapsidated conformation
of bacteriophage T7 DNA, {\it Cell}, {\bf 91}, 271-280.

\bibitem{doi} Doi M., and Edwards S.~F., (1988), {\it The theory of polymer
dynamics}, Oxford University Press, Oxford.

\bibitem{evanska} Evans E. and Skalak R.,  (1980), Mechanics and
thermodynamics of biomembranes, {\it CRC reviews in Bioeng.}, CRC Press,
Boca Raton, Florida, 1-254.

\bibitem{earnshaw} Earnshaw W.~C. and Casjens S.~R. (1980), DNA packaging
by double-stranded DNA bacteriophages, {\it Cell}, {\bf 21}, 319-331.



\bibitem{earnshaw2} Earnshaw W.~C. and Harrison S.~C. (1977), DNA

arrangement in isometric phage heads, {\it Nature}, {\bf 268}, 598-602.

\bibitem{alex} Evilevitch A., Lavelle L., Knobler C.~M., Raspaud E. and
Gelbart W.~G., (2003), Osmotic pressure inhibition of DNA ejection from
phage, {\it Proc. Natl. Acad. Sci.}, in press.

\bibitem{freusure} Freund L.~B. and Suresh S. (2003), {\it Thin film materials:
Stress, defect formation and surface evolution}, Cambridge University Press,
Cambridge.

\bibitem{goldpow} Goldstein R.~E., Powers T.~R. and Wiggins C.~H., (1998)
Viscous non-linear dynamics of twist and writhe, {\it Phys. Rev. Lett.},
{\bf 80}, 5232-5235.

\bibitem{kanamaru} Kanamaru S., Leiman P.~G., Kostyuchenko V.~A., Chipman P.~R.,
Mesyanzhinov V.~M., Arisaka F. and Rossmann M.~G., (2002), Structure of the
cell-puncturing device of bacteriophage T4, {\it Nature}, {\bf 415}, 553.

\bibitem{kindt} Kindt, J.T., Tzlil, S., Ben-Shaul, A., and Gelbart, W. (2001),
DNA packaging and ejection forces in bacteriophage, {\it Proc. Nat. Acad. Sci.},
{\bf 98}, 13671-13674.

\bibitem{lodembo} Lo C.~M., Wang H.~B., Dembo M., Wang Y.~L. (2000),
Cell movement is guided by the rigidity of the substrate, {\it Biophys. J.}
{\bf 79}, 144-152.

\bibitem{marcm} Meyers M.~A., (1994), {\it Dynamic Behavior of Materials},
John Wiley \& Sons.

\bibitem{odijk} Odijk, T., (1998), Hexagonally packed DNA within bacteriophage
T7 stabilised by curvature stress,  {\it Biophys. J.} {\bf 75}, 1223-1227 .

\bibitem{parseg} Parsegian V.~A., Rand R.~P., Fuller N.~L., and Rau D.~C.,
(1986), Osmotic stress for the direct measurement of intermolecular forces,
{\it Methods in Enzymology}, {\bf 127}, 400-416.

\bibitem{ptashne} Ptashne, M., (1992), {\it A genetic switch}, Blackwell
Science (UK).

\bibitem{purohit} Purohit P.~K., Kondev J., and Phillips R., (2003), Mechanics
of DNA packaging in viruses, {\it Proc. Nat. Acad. Sci.}, {\bf 100}(6),
3173-3178.

\bibitem{raspau} Raspaud E., de la Cruz M.~O., Sikorav J.~L., Livolant F., (1998),
Precipitation of DNA by polyamines: A polyelectrolyte behavior, {\it Biophys. J.},
{\bf 74}(1), 381-393.

\bibitem{raup} Rau, D.~C., Lee, B., and Parsegian, V.~A., (1984), Measurement
of the repulsive force between the poly-electrolyte molecules in ionic solution
- hydration forces between parallel DNA helices, {\it  Proc. Natl. Acad.  Sci.},
{\bf 81}, 2621-2625.

\bibitem{parsegian} Rau, D.~C. and Parsegian, V.~A.,(1992), Direct measurement
of the intermolecular forces between counterion-condensed DNA helices - evidence
of long-range attractive hydration forces, {\it Biophys. J.} {\bf 61 }, 246-259.

\bibitem{riemer} Riemer, S.~C. and Bloomfield, V.~A., (1978), Packaging
of DNA in bacteriophage heads: Some considerations on energetics, {\it Biopolymers}
{\bf 17}, 785-794.

\bibitem{rice} Rice J.~R., Lapusta N. and Ranjith K., (2001), Rate and
state dependent friction and the stability of sliding between elastically
deformable solids, {\it J. Mech. Phys. Sol.}, {\bf 49}(9), 1865-1898.

\bibitem{rosakis} Rosakis A.~J. (2002) Intersonic shear cracks and fault
ruptures, {\it Adv. Phys.}, {\bf 51}(4), 1189-1257.

\bibitem{smith} Smith D.~E., Tans S.~J., Smith S.~B., Grimes S., Anderson D.~L.
and Bustamante C., (2001), The bacteriophage $\phi$29 portal motor
can package DNA against a large internal force, {\it Nature}, {\bf 413}, 748.

\bibitem{simpson} Simpson A.~A., Tao Y., Leiman P.~G., Badasso M.~O., He Y.,
Jardine P.~J., Olson N.~H., Morais M.~C., Grimes S., Anderson D.~L., Baker T.~S.
and Rossmann M.~G., (2000), Structure of the bacteriophage $\phi$29 DNA packaging
motor, {\it Nature}, {\bf 408}, 745.

\bibitem{seifert} Seifert U., (1997), Configurations of fluid membranes and
vesicles, {\it Adv. Phys.}, {\bf 46}, 13.

\bibitem{tao} Tao, Y., Olson N.~H., Xu, W., Anderson, D.~L., Rossman, M.~G.,
and Baker, T.~S., (1998), Assembly of tailed bacterial virus and its genome
release studied in three dimensions, {\it Cell}, {\bf 95}, 431-437.

\bibitem{vogel} Vogel S., (1988), {\it Life's Devices: The physical world of
plants and animals}, Princeton University press, Princeton, New Jersey.

\bibitem{wang} Wang M.~D., Schnitzer S.~J, Yin H., Landick R., Gelles J. and
Block S.~M. (1998), Force and velocity measured for single molecules of
RNA polymerase, {\it Science}, {\bf 282}, 902-907.

\bibitem{widom} Widom J., (2001), Role of DNA sequence in nucleosome stability
and dynamics, {\it Q. Rev. of Biophys.}, {\bf 34}, 269-324.

\bibitem{wikoff} Wikoff W.~R. and Johnson J.~E., (1999), Imaging a molecular
machine, {\it Curr. Bio.}, {\bf 9}(8), R296-R300.

\end{thebibliography}
\end{document}